\documentclass[12pt,epsf,amssymb]{article} \textheight =23 truecm
\def\lsim{\raise0.3ex\hbox{$<$\kern-0.75em\raise-1.1ex\hbox{$\sim$}}}
\def\gsim{\raise0.3ex\hbox{$>$\kern-0.75em\raise-1.1ex\hbox{$\sim$}}}
\def\noi{\noindent} \def\nn{\nonumber} \def\bea{\begin{eqnarray}}
\def\eea{\end{eqnarray}} \def\beq{\begin{equation}}
\def\eeq{\end{equation}} 
\def\beeq{\begin{eqnarray}} \def\eeeq{\end{eqnarray}} \def\R{ {\rm R
\kern -.31cm I \kern .15cm}} \def\C{ {\rm C \kern -.15cm \vrule
width.5pt \kern .12cm}} \def\Z{ {\rm Z \kern -.27cm \angle \kern
.02cm}} \def\N{ {\rm N \kern -.26cm \vrule width.4pt \kern .10cm}}
\def\1{{\rm 1\mskip-4.5mu l} }

\usepackage{graphics} \usepackage{graphicx} \usepackage{epsfig}
\usepackage{color} \usepackage{amsmath} \usepackage[T1]{}

\begin{document} \begin{center} {\large{\bf New results on the baryon decay $\Lambda_b \to \Lambda_c \ell \bar{\nu}_\ell$}\\
\large{\bf in Heavy Quark Effective Theory}} \\

\vskip 0.3 truecm {\bf F. Jugeau$^{a,b}$, Yu Jia$^b$}\\

{\it $^a$ Instituto de F\'isica, Universidade Federal do Rio de Janeiro}\\
{\it Caixa Postal 68528, 21941-972 Rio de Janeiro, RJ, Brasil}\\
\vspace*{10pt}
{\it $^b$ Institute of High Energy Physics, Chinese Academy of Sciences}\\
{\it Theoretical Physics Center for Science Facilities}\\
{\it 19B YuQuan Lu, Shijingshan district, 100049 Beijing, China} 

\vskip 0.3 truecm {\bf L. Oliver}\\
{\it Laboratoire de Physique Th\'eorique}\footnote{Unit\'e Mixte de
Recherche UMR 8627 - CNRS }\\    {\it Universit\'e de Paris XI,
B\^atiment 210, 91405 Orsay Cedex, France} \end{center}

\vskip 0.3 truecm

\begin{abstract}
The baryon differential spectrum of the baryon decay $\Lambda_b \to \Lambda_c \ell \bar{\nu}_\ell$ will be measured in detail at LHCb. We obtain new results on the form factors in the heavy quark expansion of Heavy Quark Effective Theory that can be useful in the interpretation of the data. We formulate a sum rule for the elastic subleading form factor $A(w)$ at order $1/m_Q$, that originates from the Lagrangian perturbation $\mathcal{L}_{kin}$. In the sum rule appear only the intermediate states $\left (j^P, J^P \right ) = \left (0^+, {1 \over 2}^+ \right)$, entering also in the $1/{m_Q^2}$ correction to the axial form factor $G_1(w)$, that contributes to the differential rate at zero recoil $w = 1$. This result, together with another sum rule in the forward direction for $|G_1(1)|^2$, allows us to obtain a lower bound for the correction at zero recoil $- \delta_{1/{m_Q^2}}^{(G_1)}$ in terms of the derivative $A'_1(1)$ and the slope $\rho^2_\Lambda$ and curvature $\sigma^2_\Lambda$ of the elastic Isgur-Wise function $\xi_{\Lambda}(w)$. Another theoretical implication is that $A'(1)$ must vanish for some relation between $\rho^2_\Lambda$ and $\sigma^2_\Lambda$, as well as for $\rho^2_\Lambda \to 0$, establishing a non-trivial correlation between the leading IW function $\xi_{\Lambda}(w)$ and the subleading one $A(w)$. A phenomenological estimation of these two functions allows to obtain a lower bound on $- \delta_{1/{m_Q^2}}^{(G_1)}$.
\end{abstract}


\noi LPT Orsay 11-93 \qquad \qquad January 2012

\par 


\noindent e-mails :  jugeau@ihep.ac.cn, jiay@ihep.ac.cn, oliver@th.u-psud.fr 

\newpage\pagestyle{plain}

\section{Introduction} \hspace*{\parindent}

The present paper concerns the application, within Heavy Quark Effective Theory (HQET), of the formalism used to study the decay $\bar{B} \to D^{(*)}\ell \bar{\nu}_{\ell}$ \cite{JLOR-1} for the subleading corrections in $1/m_Q$ due to the Lagrangian perturbation in the baryon transition $\Lambda_b \to \Lambda_c \ell \bar{\nu}_{\ell}$. In the present paper, as in \cite{JLOR-1}, one important ingredient is the consideration of the non-forward amplitude $\Lambda_b(v_i) \to \Lambda_c(v') \to  \Lambda_b(v_f)$, allowing for general four-velocities $v_i$, $v_f$, $v'$, first proposed by Uraltsev in the meson case at leading order of the heavy mass expansion \cite{URALTSEV}.\par
In the meson case, at leading order, the Bjorken Sum Rule (SR) \cite{BJORKEN}\cite{IW} gives the lower bound for the derivative of the meson elastic IW function at zero recoil $w = 1$, $\rho^2 \equiv - \xi ' (1) \geq {1\over 4}$. In \cite{URALTSEV}, Uraltsev used the non-forward amplitude to derive a new SR that, combined with Bjorken's, gave the much stronger lower bound $\rho^2 \geq {3 \over 4}$. In \cite{LOR-1}, we did develop a manifestly covariant formalism within the Operator Product Expansion (OPE) and the non-forward amplitude, using the whole tower of heavy meson states \cite{FALK}. We did recover Bjorken and Uraltsev SR plus a general class of SR that also allow us to bound higher derivatives of the IW function \cite{LOR-2}. In particular, we found a bound on the curvature in terms of the slope $\rho^2$, namely $\sigma^2 \equiv \xi '' (1) \geq {1 \over 5} \left [ 4 \rho^2 + 3(\rho^2)^2 \right ]$ \cite{LOR-3}.\par
Recently, at leading order, we have extended the SR method to the elastic leading IW function $\xi_{\Lambda}(w)$ of the baryonic transition $\Lambda_b \to \Lambda_c \ell \bar{\nu}_{\ell}$ \cite{LOR-4}. We have recovered the lower bound for the slope $\rho_\Lambda^2 \equiv  - \xi '_\Lambda (1) \geq 0$ \cite{IWY} and generalized it, as we briefly summarize in Section 3.\par 
The shape of the differential spectrum of the decay $\Lambda_b \to \Lambda_c \ell \bar{\nu}_{\ell}$ depends both on these results at leading order, as well as on the $O(1/m_Q)$ corrections, which are the main subject of the present paper. The mode $\Lambda_b \to \Lambda_c \ell \bar{\nu}_{\ell}$ has a large branching ratio already measured at the Tevatron, of about 5 \%, a large fraction of the inclusive semileptonic decay $BR(\Lambda_b \to \Lambda_c \ell \bar{\nu}_{\ell} + \hbox{ anything}) \cong 10\%$. At LHCb, precise data on the decay $\Lambda_b \to \Lambda_c \ell \bar{\nu}_{\ell}$, in particular on the differential rate, will be obtained in near future.\par

The paper is organized as follows. Section 2 makes explicit the notations on the form factors and the heavy quark expansion. In Section 3, we summarize results on the leading IW function $\xi_\Lambda(w)$ for baryons. Section 4 gives a short derivation
of the relevant SR for the Lagrangian elastic $1/m_Q$ axial form factor $A(w)$ in the transition $\Lambda_b \to \Lambda_c$, by starting from the definition of the subleading Lagrangian form factors. In Section 5, we deduce another SR for the $O(1/m_Q^2)$ correction at zero recoil of the axial form factor $G_1(w)$, denoted $-\delta_{1/m^2_{Q}}^{(G_1)}$, that contributes to the differential rate at $w = 1$. We show that the same states with light cloud $j^P = 0^+$ contributing to this latter SR, contribute also to another SR for $A(w)$. In Section 6, using Schwarz inequality, we obtain a lower bound on the correction $-\delta_{1/m^2_{Q}}^{(G_1)}$ in terms of the subleading elastic form factor $A(w)$ and the leading elastic IW function $\xi_\Lambda (w)$. In Section 7, we establish a correlation between $A'(1)$ and the shape of the leading IW function and in Section 8 we summarize the physical parameters involved in the bound. In Section 9, following Leibovich and Stewart \cite{LS}, we bound one of these parameters. In Section 10, we summarize theoretical results on leading and subleading IW functions that follow from HQET and from QCD Sum Rules (QCDSR). In Section 11, we perform a calculation of $A'(1)$ in the quark model. Finally, in Section 12, we give our numerical results on the lower bound on $-\delta_{1/m^2_{Q}}^{(G_1)}$, and in Section 13, we conclude.

\vskip 1 truecm

\section{Baryon leading and subleading IW functions}

We start defining the notations for the matrix elements and the form factors following Falk and Neubert \cite{FN1r} :
\begin{align}
\langle\Lambda_c(v',s')\mid\bar{c}\gamma^{\mu}b\mid\Lambda_b(v,s)\rangle&=\bar{u}_{\Lambda_c}(v',s')\left[F_1\gamma^\mu + F_2 v^\mu + F_3 v'^\mu \right ] u_{\Lambda_b}(v,s)\;,\label{1.1e}\\
\langle\Lambda_c(v',s')\mid\bar{c}\gamma^{\mu}\gamma_5 b\mid\Lambda_b(v,s)\rangle&=\bar{u}_{\Lambda_c}(v',s')\left [G_1\gamma^\mu + G_2 v^\mu + G_3 v'^\mu \right ]\gamma_5 u_{\Lambda_b}(v,s)\label{1.2e}
\end{align}
where $u_\Lambda(v,s)$ are spinors normalized by $\bar{u}_\Lambda(v,s)u_\Lambda(v,s') = 2m_\Lambda\delta_{s,s'}$.\par

In the heavy quark mass expansion, up to the order $1/m_Q$, these form factors are given in terms of two functions $B_1(w)$ and $B_2(w)$ of $w = v\cdot v'$ \cite{GGWr} :
\begin{align}
F_1(w)&= \xi_\Lambda(w) + \left ({1 \over {2m_b}} + {1 \over {2m_c}} \right ) [B_1(w)-B_2(w)]\;,\label{1.3e}\\
G_1(w)&= \xi_\Lambda(w) + \left ({1 \over {2m_b}} + {1 \over {2m_c}} \right ) B_1(w)\;,\label{1.4e}
\end{align}
\begin{align}
F_2(w)&= G_2(w) = {1 \over {2m_c}} B_2(w)\;,\label{1.5e}\\
F_3(w)&= - G_3(w) = {1 \over {2m_b}} B_2(w)\label{1.6e}
\end{align}
where
\begin{align}
B_1(w)& = {w-1 \over w+1}\ \bar{\Lambda}\ \xi_\Lambda(w) + A(w)\;,\label{1.7e}\\
B_2(w)& = - {2 \over w+1}\ \bar{\Lambda}\ \xi_\Lambda(w)\;\label{1.8e}\;. 
\end{align}

For a generic heavy quark current $J = \bar{h}_{v'} \Gamma h_v$ ($\Gamma$ is any Dirac matrix), the elastic leading IW function $\xi_\Lambda(w)$ is defined by:
\bea
\label{1.9e}
\langle\Lambda_c(v',s')\mid\bar{h}_{v'}^{(c)}\Gamma h_v^{(b)}\mid\Lambda_b(v,s)\rangle= \xi_\Lambda(w) \bar{u}_{\Lambda_c}(v',s')\Gamma u_{\Lambda_b}(v,s)\;.
\eea

The terms proportional to $\bar{\Lambda}\xi_\Lambda(w)$ come from the $1/m_Q$ {\it Current-type perturbations}. Note that the situation for mesons is different. In this case, there are two types of current perturbations, namely, one proportional to $\bar{\Lambda}\xi(w)$ and another independent function $\xi_3(w)$ \cite{FN2r}.\par

The function $A(w)$ comes from the kinetic part $\mathcal{L}_{kin,v}$ of the Lagrangian perturbation:
\bea
\label{1.10pree}
\mathcal{L}_{eff} = \mathcal{L}_0 + \mathcal{L}_{kin,v} + \mathcal{L}_{mag,v} + O(1/m_Q^2)\;,
\eea
\bea
\label{1.10e}
\mathcal{L}_{kin,v} = {1 \over 2m_Q}\mathcal{O}^{(Q)}_{kin,v} \qquad , \qquad \mathcal{O}^{(Q)}_{kin,v} = \bar{h}_v^{(Q)} (iD_{\perp})^2 h_v^{(Q)}\;,
\eea
\bea
\label{1.10poste}
\mathcal{L}_{mag,v} = {1 \over 2m_Q}\mathcal{O}^{(Q)}_{mag,v} \qquad , \qquad \mathcal{O}^{(Q)}_{mag,v} = - {i \over 2} \bar{h}_v^{(Q)}\sigma_{\mu\nu}G^{\mu\nu} h_v^{(Q)}\;.
\eea

For a current $J = \bar{h}_{v'} \Gamma h_v$, one has: 
\bea
\label{1.11e}
\langle\Lambda_c(v',s')\mid i \int d^4xT \{ J(0), \mathcal{O}^{(c)}_{kin}(x)\}\mid\Lambda_b(v,s)\rangle = A(w)\bar{u}_{\Lambda_c} \Gamma u_{\Lambda_b}\;.
\eea

Notice that in the transition $\Lambda_b \to \Lambda_c \ell \bar{\nu}_\ell$ under consideration, the magnetic part $\mathcal{L}_{mag,v}$ of the $1/m_Q$ perturbation to the Lagrangian does not contribute \cite{GGWr}. This is unlike the case of the meson ground state $j^P = {1 \over 2}^-$ which involves instead three independent Lagrangian perturbation functions, coming from both $\mathcal{L}_{kin,v}$ and $\mathcal{L}_{mag,v}$ and denoted by $\chi_{i}(w)$ (i = 1,2,3) \cite{FN2r}.\par

An important physical remark must be emphasized here, namely, the simplicity of the baryon transition $\Lambda_b \to \Lambda_c \ell \bar{\nu}_\ell$ concerning the $1/m_Q$ corrections. Indeed, we have a single {\it Current}-type perturbation proportional to  $\bar{\Lambda}\xi_\Lambda(w)$ and a single {\it Lagrangian}-type form factor $A(w)$ originating from the kinetic part $\mathcal{L}_{kin,v}$. This simplification compared to the meson case comes merely from the fact that the light cloud in the baryon ground state $\Lambda_Q$ has $j^P = 0^+$.  This is in sharp contrast with the the meson ground state decay $B \to D(D^*) \ell \bar{\nu}_\ell$, where the light cloud has $j^P = {1 \over 2}^-$.\par 
The subleading baryon form factor $A(w)$ coming from $\mathcal{L}_{kin,v}$ satisfies, due to vector current conservation, the condition at zero recoil \cite{FN1r}:
\bea
\label{1.10bise}
A(1) = 0\;.
\eea

The differential rate for $\Lambda_b \to \Lambda_c \ell \bar{\nu}_\ell$ in the neighborhood of the zero recoil point $w = 1$ depends only on the form factor $G_1(w)$. One gets, in the limit $w \to 1$:
\bea
\label{1.12e} 
{1 \over \sqrt{w^2-1}} {d\Gamma(\Lambda_b \to \Lambda_c \ell \bar{\nu}_\ell) \over dw} \simeq {G_F^2|V_{cb}|^2 \over 4 \pi^3} m^3_{\Lambda_c} (m_{\Lambda_b}-m_{\Lambda_c})^2 |G_1(1)|^2 \qquad (w \simeq 1) 
\eea
where $G_1(1)$ has only corrections to order $1/m_Q^2$ as made explicit in formula (5.2) of \cite{FN1r}:
\bea
\label{1.13e} 
G_1(1) = 1 + \delta_{1/{m_Q^2}}^{(G_1)}\;.  
\eea
We will come back to the detail of these corrections in Section 5.

\section{Results on the leading IW function $\xi_\Lambda(w)$}

At leading order, we have extended the SR method to the baryon IW function $\xi_{\Lambda}(w)$ of the transition $\Lambda_b \to \Lambda_c \ell \bar{\nu}_{\ell}$ \cite{LOR-4}. Defining the slope and the curvature of the elastic IW function $\xi_\Lambda (w)$ as
\bea
\label{3.1e} 
\xi_\Lambda (w) = 1 - \rho_\Lambda^2 (w-1) + {\sigma_\Lambda^2 \over 2} (w-1)^2 + ...\;,
\eea
we recovered the Bjorken SR for this transition \cite{IWY}:
\beq
\label{3.2e}
\rho^2_\Lambda \equiv - \xi '_\Lambda (1) = \sum_{n\geq 0} \left [ \tau_1^{(n)} (1) \right ]^2\;.
\eeq
\noi The quantities $\tau_1^{(n)} (1) $ denote the $j^P = 0^+ \to 1^-$ IW functions at zero recoil (the light cloud has quantum numbers $j^P$ and $n$ is the radial quantum number). Let us point out that only the intermediate states $\Lambda_c^{(n)}$ with isospin $I=0$ can contribute to the SR.
\noindent Therefore, Eqn. (\ref{3.2e}) implies the lower bound for the slope \cite{IWY}:
\bea
\label{3.3e} 
\rho_\Lambda^2  \geq 0\;.
\eea 

Using the whole set of SR obtained within the non-forward amplitude method, we also obtained a new lower bound on the curvature $\sigma_\Lambda^2$:
\bea
\label{3.4e} 
\sigma_\Lambda^2 \geq {3 \over 5}\ [\rho_\Lambda^2+(\rho_\Lambda^2)^2]\;.
\eea

\noindent The bound (\ref{3.4e}) arises from the SR deduced in \cite{LOR-4}:
\beq
\label{3.5e}
 \xi ''_\Lambda (1) = 2 \sum_{n\geq 0}  \left [ \tau_{2}^{(n)}(1) \right ]^2\;,
\eeq
\beq
\label{3.6e}
2 \sum_{n\geq 0}  \left [ \tau_{2}^{(n)}(1) \right ]^2 + \sum_{n\geq 0}  \tau_{1}^{(n)}(1)\tau_{1}^{(n)}{'}(1) = 0\;,
\eeq
\begin{eqnarray}
&\displaystyle{\sum_{n\geq 0}  \left [ \tau_{1}^{(n)}(1) \right ]^2 + {8 \over 3} \sum_{n\geq 0}  \left [ \tau_{2}^{(n)}(1) \right ]^2 + \sum_{n\geq 0}  \left [  \xi ^{(n)}_\Lambda{'} (1)\right ]^2}& \nn \\
&\displaystyle{+4  \sum_{n\geq 0}  \tau_{1}^{(n)}(1)\tau_{1}^{(n)}{'}(1) +  \xi ''_\Lambda (1) = 1\;.}&\label{3.7e}
\end{eqnarray}
\noindent Eliminating the unknown quantity $\sum\limits_{n\geq 0}  \tau_{1}^{(n)}(1)\tau_{1}^{(n)}{'}(1)$, we finally obtained for the curvature:
\beq
\label{3.8e}
\sigma_\Lambda^2 \equiv \xi ''_\Lambda (1) = {3 \over 5} \left \{  \rho_\Lambda^2 + (\rho_\Lambda^2)^2 + \sum_{n\not= 0} \left [  \xi^{(n)}{'} (1) \right ]^2 \right \}
\eeq

\noi that implies the improved bound (\ref{3.4e}).

In \cite{LOR-5}, a group theoretical method to study IW functions has been exposed, that sheds a new light on these results. 
In this approach, a current matrix element splits into a heavy quark matrix element and an overlap of the initial and final clouds, related to the IW functions, that contain the long distance physics \cite{FALK}. The light cloud belongs to the Hilbert space of a unitary representation of the Lorentz group. Decomposing into irreducible representations, one obtains the IW function as an integral formula, superposition of {\it irreducible} IW functions with {\it positive measures}, providing positivity bounds on their derivatives.\par
One demonstrated in \cite{LOR-5} that this Lorentz group method is equivalent to the SR approach, and summarized all the possible constraints. The general formalism was thoroughly applied to the case $j = 0$ for the light cloud, relevant to $\Lambda_b \to \Lambda_c \ell \bar{\nu}_{\ell}$. One recovers the bounds (\ref{3.3e}) and (\ref{3.4e}) and gets new bounds for higher derivatives. Also, the Lorentz group approach provides rigorous criteria to decide if a given ansatz for the IW function is compatible or not with the general SR, and therefore with HQET. \par
Let us now recall a result that will become relevant in the present paper. In \cite{LOR-5}, it has been demonstrated that the $k$-th derivative of the elastic IW function is given by the expectation value of a polynomial of degree $k$:
\beq
\label{3.9e}
\xi_\Lambda^{(k)}(1) = (-1)^k\ 2^k {k! \over (2k+1)!} <\prod^k_{i=1} (x+i^2-1)>
\eeq
with the expectation value defined by
\beq
\label{3.10e}
< f(x) >\ = \int_0^\infty f(x)\ d\nu(x)
\eeq
where $\nu$ is a normalized positive measure with support in $[0,\infty[$.\par
One obtains for the first two derivatives: 
$$\rho_\Lambda^2 = {1 \over 3} <x>\;,$$
\beq
\label{3.11e}
\sigma_\Lambda^2 = {1 \over 15} <x(x+3)>\;.
\eeq

The point is the following. If the first derivative $\rho_\Lambda^2$ attains its lowest possible value (\ref{3.3e}), then one gets for the first moment:
\beq
\label{3.12e}
<x>\ = \int_0^\infty x\ d\nu(x) = 0
\eeq
that completely determines the measure $\nu$:
\beq
\label{3.13e}
<x>\ = 0 \qquad\qquad \Leftrightarrow \qquad\qquad d\nu(x) = \delta(x)\ dx\;.
\eeq
\noindent This implies, for the different moments: 
\beq
\label{3.14e}
<x^k>\ = 0
\eeq
\noindent and therefore
\beq
\label{3.15e}
\rho_\Lambda^2\ = 0 \qquad\qquad \Rightarrow \qquad\qquad \xi_\Lambda^{(k)}(1) = 0\;.
\eeq 
\noindent As a particular case one gets the result concerning the curvature:
\beq
\label{3.16e}
\rho_\Lambda^2\ = 0 \qquad\qquad \Rightarrow \qquad\qquad \sigma_\Lambda^2\ = 0
\eeq

\noindent that we will use below. 

\section{SR for the subleading $1/m_Q$ form factor $A(w)$} \hspace*{\parindent}

From the definition (\ref{1.11e}), one gets, inserting intermediate states and taking into account that $\mathcal{L}_{mag}$ does not contribute:
\bea
\label{4.1e} 
A(w) = \sum_{n\not= 0} {1 \over \Delta E^{(n)}} \xi^{(n)}_\Lambda(w)\frac{\langle\Lambda_c^{(n)}(v,s)\mid\mathcal{O}_{kin,v}^{(c)} (0)\mid\Lambda_c(v,s)\rangle}{\sqrt{4m_{\Lambda_c^{(n)}}m_{_{\scriptstyle\Lambda_c}}}\sqrt{v^0_{\Lambda_c^{(n)}}v^0_{_{\scriptstyle\Lambda_c}}}}
\eea
where $\Delta E^{(n)} \equiv m_{\Lambda_c^{(n)}} - m_{_{\scriptstyle\Lambda_c}}$ is the mass difference. Since, as argued above, only the operator $O_{kin,v}^{(c)}$ contributes in this relation and $\Lambda_c^{(n)}$ is an excited state with the same quantum numbers $J^P_j = {1 \over 2}^+_0$ as the ground state. The only possibility is for $j = L = 0$, which means that in the sum (\ref{4.1e}) only the radial excitations contribute.\par

From (\ref{4.1e}), we realize that, as it should, $A(1) = 0$ because $\xi^{(n)}_\Lambda(1) = \delta_{n,0}$ and
the slope of $A(w)$ at $w = 1$ is given by the expression 
\bea
\label{4.2e} 
A'(1) = \sum_{n\not= 0} {1 \over \Delta E^{(n)}} \xi_\Lambda^{(n)'}(1) {\langle\Lambda_c^{(n)}(v,s)\mid\mathcal{O}_{kin,v}^{(c)} (0)\mid\Lambda_c(v,s)\rangle \over \sqrt{4m_{\Lambda_c^{(n)}}m_{_{\scriptstyle\Lambda_c}}}\sqrt{v^0_{\Lambda_c^{(n)}}v^0_{_{\scriptstyle\Lambda_c}}}}\;.
\eea

\noindent This relation will be used below to obtain a bound on the $O(1/m_Q^2)$ correction at zero recoil of the axial form factor $G_1(1)$.

\vskip 1 truecm

\section{Sum rule for the $\delta_{1/{m_Q^2}}^{(G_1)}$ correction} \hspace*{\parindent}

Considering the spatial component of the {\it axial-vector current} and keeping terms of $O(1/m_Q^2)$, one can write a SR similar to the one formulated in the meson case (formulas (114) of \cite{BIGI} and (5.6) of \cite{LLSW}):
\begin{eqnarray} 
&\displaystyle\mid G_1(1)\mid^2 + {1 \over 2}\sum_{s,s'} \sum_{n\not= 0}{\mid\langle \Lambda_c^{(n)}(0^+,1^+)(v,s')\mid \vec{A}\mid\Lambda_b(v,s) \rangle\mid^2 \over 4m_{\Lambda_c^{(n)}}m_{_{\scriptstyle\Lambda_b}}}&\nn\\
&\displaystyle = \eta_A^2 + \left [ \left ({1 \over 2m_c}  - {1 \over 2m_b} \right )^2 + {8 \over 3} {1 \over 2m_c} {1 \over 2m_b} \right ] \lambda & \label{5.1e} 
\end{eqnarray}
where the parameter $- \lambda$ is the mean kinetic energy value defined by
\bea
\label{5.2e} 
- \lambda = \mu_\pi^2 = {1 \over 2m_{\Lambda_b}} \langle\Lambda_b(v)\mid\bar{h}_v^{(b)}(iD)^2 h_v^{(b)}\mid\Lambda_b(v)\rangle\;.
\eea
The factor ${1 \over 2}$ in front of the sum (\ref{5.1e}) comes from average over the spins $s$ of the initial state, and is consistent with the appearance of $|G_1(1)|^2$ in the \emph{l.h.s.} and also with the OPE in the \emph{r.h.s.}. The \emph{r.h.s.} of (\ref{5.1e}) is reminiscent of the one found in the meson case with the difference that the magnetic part vanishes for the $\Lambda_Q$ ground state $\frac{1}{2}^+$. Notice that in formula \eqref{5.1e} and in what follows, a single spatial component of the axial-vector current should be understood to contribute.\par
A similar SR has also been obtained in the baryon case for the  {\it vector current} and with intermediate states $j^P = 1^-$ ($J^P = {1 \over 2}^-, {3 \over 2}^-$) \cite{LS}.\par
  
In formula (\ref{5.1e}), one must keep in mind that the matrix element $< \Lambda_c^{(n)} |\vec{A}| \Lambda_b >$ implicitly contains the double insertions of the kinetic and magnetic parts of the Lagrangian perturbations to the axial-vector current $A^\mu = \overline{c} \gamma^\mu \gamma_5 b$. Only the spatial part $\vec{A} = \overline{c} \vec{\gamma} \gamma_5 b$ contributes since in the heavy quark limit the component $\overline{c} \gamma^0 \gamma_5 b$ is a subleading corrections to the current itself. Also, in the sum of the \emph{l.h.s.} of (\ref{5.1e}), only the states with the quantum numbers $\Lambda_c^{(n)}(j^P=0^+,1^+)$ contribute, as will become clear below.\par

Writing $G_1(1)$ under the form:
\bea
\label{5.3e} 
 G_1(1) = \eta_A + \delta_{1/{m_Q^2}}^{(G_1)}\;,
\eea

\noindent we get  the following expression for the $1/m_Q^2$ corrections:
$$- \delta_{1/{m_Q^2}}^{(G_1)} = - {1 \over 2} \left [ \left ({1 \over 2m_c}  - {1 \over 2m_b} \right )^2 + {8 \over 3} {1 \over 2m_c} {1 \over 2m_b} \right ] \lambda$$
$$+ {1 \over 4} \sum_{s,s'}  \sum_{n\not=0}{\mid\langle \Lambda_c^{(n)}(0^+)(v,s') \mid\vec{A}\mid \Lambda_b(v,s) \rangle\mid^2 \over 4m_{\Lambda_c^{(n)}}m_{_{\scriptstyle\Lambda_b}}} $$
\bea
\label{5.4e} 
+ {1 \over 4}  \sum_{s,s'} \sum_{n\not= 0}{\mid\langle \Lambda_c^{(n)}(1^+)(v,s') \mid\vec{A}\mid \Lambda_b(v) \rangle\mid^2 \over 4m_{\Lambda_c^{(n)}}m_{_{\scriptstyle\Lambda_b}}}\;. 
\eea

We now consider separately the final states $\Lambda_c^{(n)}(0^+,1^+)$ i.e. with quantum numbers $j^P = 0^+, J^P = {1 \over 2}^+$ that are attained by $\mathcal{L}_{kin,v}$ insertions, and $j^P = 1^+, J^P = {1 \over 2}^+ \rm{and}\ {3 \over 2}^+$ that are coupled to $\mathcal{L}_{mag,v}$ insertions.

\subsection{Matrix elements $\langle\Lambda_c^{(n)} |\vec{A}| \Lambda_b\rangle$ of transitions $0^+ \to 0^+$} 

Let us now compute the contribution 
\bea
\label{5.4bise} 
 \sum_{s,s'}\sum_{n\not= 0} {\mid\langle \Lambda_c^{(n)}(0^+)(v,s') \mid\vec{A}\mid \Lambda_b(v,s) \rangle\mid^2 \over 4m_{\Lambda_c^{(n)}}m_{_{\scriptstyle\Lambda_b}}}
\eea
to Eqn. (\ref{5.4e}). In this relation, the matrix element $\langle\Lambda_c^{(n)}|\bar{h}_v^{(c)}\vec{A}h_v^{(b)}(0)|\Lambda_b\rangle$, that contains the $\mathcal{L}_{kin,v}$ insertions on the initial and final legs, is given by: 
$$\langle \Lambda_c^{(n)}(v,s') \mid\bar{h}_v^{(c)}\vec{A}h_v^{(b)}(0)\mid \Lambda_b(v,s) \rangle\underset{O(1/m_Q)}{=}$$ 
\begin{equation}\label{5.5e}
\frac{-1}{\Delta E^{(n)}}\left(\frac{1}{2m_c}-\frac{1}{2m_b}\right)\frac{\langle\Lambda_c^{(n)}(v,s)\mid\mathcal{O}^{(c)}_{kin,v}(0)\mid \Lambda_c(v,s)\rangle}{\sqrt{4m_{\Lambda_c^{(n)}}m_{_{\scriptstyle\Lambda_c}}}\sqrt{v^0_{\Lambda_c^{(n)}}v^0_{_{\scriptstyle{\Lambda_c}}}}}u_{\Lambda_c}^\dagger(v,s') \vec{\Sigma} u_{\Lambda_b}(v,s)\;.
\end{equation}

The proof of this last relation is as follows. To compute this quantity, we must insert the $1/m_Q$ Lagrangian perturbations on the $b$ and $c$ quark legs. Since we have seen that $\mathcal{L}_{mag,v}$ does not contribute when the light cloud has $j^P = 0^+$, we have (we make lighter the formulas by skipping the normalization of the states): 
\begin{eqnarray}
&\langle \Lambda_c^{(n)}(v,s') \mid\bar{h}_v^{(c)}\vec{A}h_v^{(b)}(0)\mid \Lambda_b(v,s)\rangle\underset{O(1/m_Q)}{=}&\nn\\ 
&{1 \over 2m_c} \langle\Lambda^{(n)}_c(v,s') \mid i\int d^4xT \{\mathcal{O}^{(c)}_{kin,v}(x),\bar{h}_v^{(c)}\vec{A}h_v^{(b)}(0)\}\mid \Lambda_b(v,s)\rangle&\nn\\
&+\ {1 \over 2m_b} \langle\Lambda^{(n)}_c(v,s') \mid i \int d^4xT \{\mathcal{O}^{(b)}_{kin,v}(x),\bar{h}_v^{(c)}\vec{A}h_v^{(b)}(0)\}\mid \Lambda_b(v,s) \rangle\;.&\label{5.6e} 
\end{eqnarray} 
\noindent Inserting intermediate states, one has:
\begin{eqnarray}
&\langle\Lambda^{(n)}_c(v,s')\mid i\int d^4xT \{\mathcal{O}^{(c)}_{kin,v}(x),\bar{h}_v^{(c)}\vec{A}h_v^{(b)}(0)\}\mid \Lambda_b(v,s)\rangle=&\nn\\
&-{1 \over \Delta E^{(n)}}\langle\Lambda_c^{(n)}(v,s')\mid\mathcal{O}^{(c)}_{kin,v}(0)\mid \Lambda_c(v,s')\rangle\langle\Lambda_c(v,s')\mid\bar{h}_v^{(c)}\vec{A}h_v^{(b)}(0)\mid \Lambda_b(v,s)\rangle & \label{5.7e} 
\end{eqnarray}
\noindent and
\begin{eqnarray}
&\langle\Lambda^{(n)}_c(v,s)	\mid i \int d^4xT \{\mathcal{O}^{(b)}_{kin,v}(x),\bar{h}_v^{(c)}\vec{A}h_v^{(b)}(0)\}\mid \Lambda_b(v,s)\rangle=&\nn\\
&+{1 \over \Delta E^{(n)}}\langle\Lambda_c^{(n)}(v,s') \mid\bar{h}_v^{(c)}\vec{A}h_v^{(b)}(0)\mid \Lambda_b^{(n)}(v,s)\rangle\langle \Lambda_b^{(n)}(v,s) \mid\mathcal{O}^{(b)}_{kin,v}(0)\mid \Lambda_b(v,s) \rangle& \label{5.8e} 
\end{eqnarray}
\noindent because $O_{kin}^{(Q)}$ conserves the spin projection. Using heavy quark flavor-spin symmetry, the fact that the matrix elements are at zero recoil and that the matrix elements of $O_{kin}^{(Q)}$ are independent of the spin projection:
\begin{align}
\langle\Lambda_b^{(n)}(v,s)\mid\mathcal{O}^{(b)}_{kin,v}(0)\mid\Lambda_b(v,s)\rangle &=\langle\Lambda_c^{(n)}(v,s)\mid\mathcal{O}^{(c)}_{kin,v}(0)\mid\Lambda_c(v,s)\rangle\;,\nn\\
\langle\Lambda_c^{(n)}(v,s')\mid\mathcal{O}^{(c)}_{kin,v}(0)\mid \Lambda_c(v,s')\rangle &=\langle \Lambda_c^{(n)}(v,s)\mid\mathcal{O}^{(c)}_{kin,v}(0)\mid \Lambda_c(v,s)\rangle\;,\nn\\
\langle \Lambda_c(v,s') \mid\bar{h}_v^{(c)}\vec{A}h_v^{(b)}(0)\mid \Lambda_b(v,s)\rangle &=\langle \Lambda_c^{(n)}(v,s') \mid\bar{h}_v^{(c)}\vec{A}h_v^{(b)}(0)\mid \Lambda_b^{(n)}(v,s) \rangle\label{5.9e} 
\end{align}
\noindent and from the relation for the free current (without $\mathcal{L}$ insertions): 
\bea
\label{5.10e} 
\langle\Lambda_c(v,s')\mid\bar{h}_v^{(c)}\vec{A}h_v^{(b)}(0)\mid \Lambda_b(v,s)\rangle\underset{O\left((1/m_Q)^0\right)}{=} u_{\Lambda_c}^\dagger(v,s') \vec{\Sigma} u_{\Lambda_b}(v,s)\;,
\eea
the formula (\ref{5.5e}) follows.\par

\subsection{Matrix elements $\langle\Lambda_c^{(n)} |\vec{A}| \Lambda_b\rangle$ of transitions $0^+ \to 1^+$}

The current $\vec{A}$ involves a change of flavor $b \to c$ and the spin matrix element $u_{\Lambda_c}^\dagger(v,s') \vec{\Sigma} u_{\Lambda_b}(v,s)$, as we have seen above. To compute the matrix elements $\langle\Lambda_c^{(n)}(1^+) |\vec{A}| \Lambda_b\rangle$, we need to take into account that the spin of the final states can be $J = {1 \over 2}$ or $J = {3 \over 2}$ and consider the sums of products of matrix elements (for example, the current $A^z$ and the spin projection $J^z = + {1 \over 2}$):
\begin{align}
&{1 \over \Delta E_1^{(n)}} {1 \over {2m_c}} \langle \Lambda_c^{(n)}(1^+)  ({\scriptstyle{1 \over 2},+ {1 \over 2}})\mid\mathcal{O}^{(c)}_{mag,v}\mid \Lambda_c(0^+) ({\scriptstyle{1 \over 2},+ {1 \over 2}})\rangle\langle \Lambda_c(0^+) ({\scriptstyle{1 \over 2},+ {1 \over 2}})\mid\Sigma^z \tau^{cb}\mid \Lambda_b(0^+) ({\scriptstyle{1 \over 2},+ {1 \over 2}})\rangle\nn\\
-&{1 \over \Delta E_1^{(n)}} {1 \over {2m_b}} \langle \Lambda_c^{(n)}(1^+) ({\scriptstyle{1 \over 2},+ {1 \over 2}})\mid\Sigma^z \tau^{cb}\mid \Lambda_b^{(n)}(1^+) ({\scriptstyle{1 \over 2},+ {1 \over 2}}) \rangle\langle \Lambda_b^{(n)}(1^+) ({\scriptstyle{1 \over 2},+ {1 \over 2}})\mid\mathcal{O}^{(b)}_{mag,v}\mid \Lambda_b(0^+) ({\scriptstyle{1 \over 2},+ {1 \over 2}}) \rangle\label{5.11e}
\end{align}
\noindent and
\begin{equation}
- {1 \over \Delta E_1^{(n)}} {1 \over {2m_b}} \langle \Lambda_c^{(n)}(1^+) ({\scriptstyle{3 \over 2},+ {1 \over 2}})\mid\Sigma^z \tau^{cb}\mid \Lambda_b^{(n)} (1^+)({\scriptstyle{1 \over 2},+ {1 \over 2}}) \rangle\langle \Lambda_b^{(n)}(1^+) ({\scriptstyle{1 \over 2},+ {1 \over 2}})\mid\mathcal{O}^{(b)}_{mag}\mid \Lambda_b(0^+) ({\scriptstyle{1 \over 2},+ {1 \over 2}}) \rangle\label{5.12e}
\end{equation}
\noi where $\Delta E_1^{(n)} = m_{\Lambda_c^{(n)}(1^+)}-m_{_{\scriptstyle\Lambda_c(0^+)}}$.
Taking into account that the current operator $\Sigma^z$ acts on the heavy quark, applying the Wigner-Eckart theorem and summing incoherently the contributions with final spin ${1 \over 2}$ and ${3 \over 2}$, we obtain an expression proportional to 
$$\sum_{\Lambda_c^{(n)}} {\mid\langle \Lambda_c^{(n)}(1^+) \mid\vec{A}\mid \Lambda_b \rangle\mid^2 \over 4m_{\Lambda_c^{(n)}}m_{_{\scriptstyle\Lambda_c}}}$$
$$$$ 
\begin{align}
\propto &\sum_n \left({\langle {1 \over 2} \mid\mid\Sigma\mid\mid {1 \over 2} \rangle \over \Delta E_1^{(n)}}\right)^2 \left[ \left({1 \over 3}{1 \over {2m_b}}  + {1 \over {2m_c}}  \right)^2 + \left({1 \over {2m_b}} {2\sqrt{2} \over 3} \right)^2 \right]\nn\\
=&\sum_n \left({\langle {1 \over 2}\mid\mid\Sigma\mid\mid {1 \over 2} \rangle \over \Delta E_1^{(n)}}\right)^2 \left[ \left({1 \over {2m_b}} - {1 \over {2m_c}} \right)^2 + {8 \over 3} {1 \over {2m_c}} {1 \over {2m_b}} \right] \geq 0\;.\label{5.13e}
\end{align}

\subsection{Comparison with the $1/m_Q^2$ corrections to $G_1(1)$}

Using formulas (4.6), (4.19) and (4.20) from ref. \cite{FN1r}, the axial-vector form factor (\ref{1.2e}) at zero recoil, $G_1(1)$, can be cast into the form:
\begin{eqnarray}
&G_1(1)=\eta_A + {1 \over 2} \left[\left({1 \over {2m_c}}-{1 \over {2m_b}}\right)^2 + {8 \over 3} {1 \over {2m_c}}{1 \over {2m_b}}\right] \lambda &\nn\\
&+{1 \over 2} \left({1 \over {2m_c}}-{1 \over {2m_b}}\right)^2 \left[-D_1(1)+3D_2(1)\right] + {1 \over {2m_c}}{1 \over {2m_b}} 4D_2(1)& \label{5.14e} 
\end{eqnarray}
where the functions $D_i$ ($i = 1, 2$) correspond to the double insertions of the operators $O^{(Q)}_{kin,v}$ (\ref{1.10e}) and $O^{(Q)}_{mag,v}$  (\ref{1.10poste}) $(Q = c, b)$ on the initial and final heavy quark legs (the notation $D_i$ stands for double insertions).\par

Taking the square of (\ref{5.14e}) and neglecting cross-terms between radiative $\eta_A$ and $1/m_Q^2$ corrections, one finds, comparing to the SR (\ref{5.1e}), the relation:
\begin{eqnarray}
&\displaystyle\sum_{\Lambda_c^{(n)}} {\mid\langle \Lambda_c^{(n)}(0^+,1^+)(v) \mid\vec{A}\mid \Lambda_b(v) \rangle\mid^2 \over 4m_{\Lambda_c^{(n)}}m_{_{\scriptstyle\Lambda_b}}}=&\nn\\ 
&\left({1 \over {2m_c}}-{1 \over {2m_b}}\right)^2 \left[D_1(1)-3D_2(1)\right] - {1 \over {2m_c}}{1 \over {2m_b}} 8D_2(1)\;.&  \label{5.15e}
\end{eqnarray}
From (\ref{5.5e}) and (\ref{5.13e}), we realize that indeed, as expected for coherence, both the \emph{l.h.s.} and \emph{r.h.s.} of (\ref{5.15e}) have the same functional dependence in ${1 \over 2m_c}$ and  ${1 \over 2m_b}$.


\section{Bound on the $O(1/m_Q^2)$ correction at zero recoil} \hspace*{\parindent}

From the positivity of (\ref{5.13e}) we obtain, from the expression (\ref{5.4e}), the inequality:
\begin{align}
- \delta_{1/{m_Q^2}}^{(G_1)} \geq & - {1 \over 2} \left [ \left ({1 \over 2m_c}  - {1 \over 2m_b} \right )^2 + {8 \over 3} {1 \over 2m_c} {1 \over 2m_b} \right ] \lambda\nn\\
&+ {1 \over 4}  \sum_{s,s'} \sum_{n\not= 0}{\mid\langle \Lambda_c^{(n)}(0^+)(v,s') \mid\vec{A}\mid \Lambda_b(v,s) \rangle\mid^2 \over 4m_{\Lambda_c^{(n)}}m_{_{\scriptstyle\Lambda_b}}} \label{6.1-1e} 
\end{align}
\noindent that gives, from (\ref{5.5e}):
\begin{align}
- \delta_{1/{m_Q^2}}^{(G_1)} \geq & - {1 \over 2} \left [ \left ({1 \over 2m_c}  - {1 \over 2m_b} \right )^2 + {8 \over 3} {1 \over 2m_c} {1 \over 2m_b} \right ] \lambda\nn\\
&+ {1 \over 2} \left({1 \over 2m_c}  - {1 \over 2m_b} \right)^2 \sum_{n\not=0} \left[{1 \over \Delta E^{(n)}} {\langle\Lambda_c^{(n)}(v,s)\mid\mathcal{O}^{(c)}_{kin,v}(0)\mid\Lambda_c(v,s)\rangle \over \sqrt{4m_{\Lambda_c^{(n)}}m_{_{\scriptstyle\Lambda_c}}}\sqrt{v^0_{\Lambda_c^{(n)}}v^0_{_{\scriptstyle\Lambda_c}}}}\right] ^2 \label{6.1-2e} 
\end{align}
\noindent where we have used the equality 
\bea
\label{6.1-3e} 
{1 \over 2}\sum_{s,s'} {\mid u_{\Lambda_c}^\dagger(v,s') \vec{\Sigma} u_{\Lambda_b}(v,s)\mid^2 \over 4m_{\Lambda_c}m_{\Lambda_b}} = 1
\eea
\noindent since it is understood that a single component of $\vec{\Sigma}$ contributes.

Using now the Schwarz inequality
\beq
\label{6.1e}
\left | \sum_n A_n B_n\right | \leq \sqrt{\left ( \sum_n |A_n|^2\right ) \left ( \sum_n |B_n|^2\right )}
\eeq
\noi to Eqn. (\ref{4.1e}):
\bea
\label{6.2e}
\left [ A(w) \right ]^2 \leq\sum_{n\not=0} \left [ \xi^{(n)}_\Lambda (w)\right ]^2 \sum_{n\not= 0}
\left [{1 \over \Delta E^{(n)}} {\langle\Lambda_c^{(n)}(v,s)\mid\mathcal{O}_{kin, v}^{(c)}(0)\mid\Lambda_c(v,s)\rangle \over \sqrt{4m_{\Lambda_c^{(n)}}m_{_{\scriptstyle\Lambda_c}}}\sqrt{v^0_{\Lambda_c^{(n)}}v^0_{_{\scriptstyle\Lambda_c}}}}\right ]^2\;,
\eea
\noi we therefore obtain:
\begin{eqnarray}
- \delta_{1/{m_Q^2}}^{(G_1)} &\geq &- {1 \over 2} \left [ \left ({1 \over 2m_c}  - {1 \over 2m_b} \right )^2 + {8 \over 3} {1 \over 2m_c} {1 \over 2m_b} \right ] \lambda\nn\\
&&\displaystyle{ +\ {1 \over 2} \left ( {1 \over 2m_c} - {1 \over 2m_b} \right )^2 \frac{[A(w)]^2}{\sum_{n\not=0} \left [ \xi^{(n)}_\Lambda (w)\right ]^2}\;.}\label{6.3e} 
\end{eqnarray}
\noindent Since the inequality (\ref{6.3e}) is valid for any value of the variable $w$, we can consider its limit for $w \to 1$, taking into account that $A(1) = 0$ and $\xi^{(n)}_\Lambda(1) = \delta_{n,0}$:
\begin{eqnarray}
- \delta_{1/{m_Q^2}}^{(G_1)} &\geq & - {1 \over 2} \left [ \left ({1 \over 2m_c}  - {1 \over 2m_b} \right )^2 + {8 \over 3} {1 \over 2m_c} {1 \over 2m_b} \right ] \lambda\nn\\
&&\displaystyle{+\ {1 \over 2} \left ( {1 \over 2m_c} - {1 \over 2m_b}\right )^2 {[A'(1)]^2 \over \sum_{n\not=0} \left[ \xi^{(n)}_\Lambda{'}(1)\right ]^2}\;.}\label{6.4e} 
\end{eqnarray}
\noindent From Eqn. (\ref{3.8e}) obtained in ref. \cite{LOR-4}:
\bea
\label{6.5e} 
\sum_{n\not=0} \left [ \xi^{(n)}_\Lambda{'}(1)\right ]^2 = {5 \over 3} \sigma_\Lambda^2-\rho_\Lambda^2-(\rho_\Lambda^2)^2\;,
\eea
one gets:
\begin{eqnarray}
- \delta_{1/{m_Q^2}}^{(G_1)}&\geq& - {1 \over 2} \left [ \left ({1 \over 2m_c}  - {1 \over 2m_b} \right )^2 + {8 \over 3} {1 \over 2m_c} {1 \over 2m_b} \right ] \lambda\nn\\
&&+\ {3 \over 10} \left ( {1 \over 2m_c} - {1 \over 2m_b}\right )^2 {[A'(1)]^2 \over \sigma_\Lambda^2 - {3 \over 5}[\rho_\Lambda^2+(\rho_\Lambda^2)^2]}\;.\label{6.6e} 
\end{eqnarray}
\vskip 0.5 truecm
This inequality is one of the main results of the present paper.\par 

From (\ref{1.12e}) and (\ref{1.13e}), we see that the $1/m_Q^2$ correction at zero recoil $- \delta_{1/{m_Q^2}}^{(G_1)}$ is crucial in the extrapolation of the differential rate $d\Gamma(\Lambda_b \to \Lambda_c \ell \overline{\nu}_\ell)/dw$ near zero recoil. This is needed to check that the value of $|V_{cb}|$ that would fit the exclusive baryon semileptonic data is indeed consistent with what we presently know on this parameter from the meson exclusive and inclusive determinations. It is in this respect that the bound (\ref{6.6e}) has to be used.\par
To get a quantitative estimate on the bound (\ref{6.6e}), we need information on the parameters $-\lambda$, $\rho_\Lambda^2$, $\sigma_\Lambda^2$ and $A'(1)$. These quantities are on quite different footing and their plausible values will be discussed in Sections 9, 10 and 11.\par

\section{Correlation between $A'(1)$ and the shape of the leading IW function $\xi_\Lambda(w)$} \hspace*{\parindent}

Since the inequality (\ref{6.6e}) holds for any values of $\rho_\Lambda^2$ and $\sigma_\Lambda^2$ satisfying the general constraints (\ref{3.3e}) and (\ref{3.4e}), it should hold, in particular, for their lowest values. Indeed, one has that if $\rho_\Lambda^2$ and $\sigma_\Lambda^2$ attain their lowest values, the \emph{r.h.s.} of the precedent inequality would diverge. This feature seems completely unphysical. Therefore, we expect a correlation between $A'(1)$ and $\sigma_\Lambda^2$, namely, one should expect that $A'(1)$ vanishes if $\sigma_\Lambda^2$ attains its lowest bound (\ref{3.4e}):
\bea
\label{7.1e} 
A'(1) \to 0 \qquad \qquad \textrm{for} \qquad \qquad \sigma_\Lambda^2 \to {3 \over 5}\ [\rho_\Lambda^2+(\rho_\Lambda^2)^2]\;.
\eea

We give now an explicit and independent demonstration of this interesting feature.\par
Let us consider the derivative of (\ref{4.1e}) at zero recoil :
\bea
\label{7.2e} 
A'(1) = \sum_{n\not= 0} {1 \over \Delta E^{(n)}} \xi_\Lambda^{(n)'}(1) {\langle\Lambda_c^{(n)}(v,s)\mid\mathcal{O}_{kin,v}^{(c)} (0)\mid\Lambda_c(v,s)\rangle \over \sqrt{4m_{\Lambda_c^{(n)}}m_{_{\scriptstyle\Lambda_c}}}\sqrt{v^0_{\Lambda_c^{(n)}}v^0_{_{\scriptstyle\Lambda_c}}}}\;.
\eea

\noindent Using again the Schwarz inequality as in Section 6, we obtain:
\beq
\label{7.3e}
[A'(1)]^2 \leq \sum_{n\not= 0} \left[\xi_\Lambda^{(n)'}(1)\right ]^2\sum_{n\not= 0} \left [{\langle\Lambda_c^{(n)}(v,s)\mid\mathcal{O}_{kin,v}^{(c)} (0)\mid\Lambda_c(v,s)\rangle \over \sqrt{4m_{\Lambda_c^{(n)}}m_{\Lambda_c}}\sqrt{v^0_{\Lambda_c^{(n)}}v^0_{_{\scriptstyle\Lambda_c}}}}\right]^2
\eeq

\noi and from relation (\ref{3.8e}) we obtain:
\beq
\label{7.4e}
[A'(1)]^2 \leq {5 \over 3} \left\{\sigma_\Lambda^2 - {3 \over 5}[\rho_\Lambda^2+(\rho_\Lambda^2)]\right \}\sum_{n\not= 0} \left [{\langle\Lambda_c^{(n)}(v,s)\mid\mathcal{O}_{kin,v}^{(c)} (0)\mid\Lambda_c(v,s)\rangle \over \sqrt{4m_{\Lambda_c^{(n)}}m_{\Lambda_c}}\sqrt{v^0_{\Lambda_c^{(n)}}v^0_{_{\scriptstyle\Lambda_c}}}}\right]^2\;.
\eeq

\noi Therefore, if the lower bound (\ref{3.4e}) is saturated, one obtains
\beq
\label{7.5e}
\sigma_\Lambda^2  = {3 \over 5}[\rho_\Lambda^2+(\rho_\Lambda^2)^2] \qquad \qquad \Rightarrow \qquad \qquad A'(1) = 0 
\eeq

\noi as we see from inspection of relation (\ref{7.4e}).\par
On the other hand, one has the important result (\ref{3.16e}), \emph{i.e.} $\rho_\Lambda^2\ = 0$ implies $\sigma_\Lambda^2\ = 0$. This feature has a clear theoretical-group interpretation \cite{LOR-5}, that implies: 

\beq
\label{7.6e}
\rho_\Lambda^2  = 0 \qquad \qquad \Rightarrow \qquad \qquad A'(1) = 0\;. 
\eeq

Formulas (\ref{7.5e}) and (\ref{7.6e}) are quite nontrivial because they imply a strong correlation between the shapes of the leading IW function $\xi_\Lambda(w)$ and of the subleading form factor $A(w)$.

\vskip 0.5 truecm

\section{Parameters involved in the inequality on $- \delta_{1/{m_Q^2}}^{(G_1)}$}

The inequality (\ref{6.6e}) involves, besides the quark masses $m_c$ and $m_b$, the  slope $\rho_\Lambda^2$ and the curvature $\sigma_\Lambda^2$ of the elastic leading IW function $\xi_\Lambda(w)$, the mean kinetic energy value $- \lambda$ (\ref{5.2e}) and $A'(1)$, the derivative of the subleading form factor $A(w)$ (\ref{1.11e}) at zero recoil. In order to compute the \emph{r.h.s.} of (\ref{6.6e}), we need an estimate of these quantities. The theoretical work on the IW function $\xi_\Lambda(w)$ and the subleading form factor $A(w)$ have not been so exhaustive as the corresponding ones in $B$ meson semileptonic decays. We will now make a review of what can be known about these parameters, namely, within HQET, the QCDSR approach and the quark model. Notice that we concentrate on the baryon $\Lambda_Q$ and leave aside other heavy baryons.\par

\section{Discussion of the parameter $-\lambda$} \hspace*{\parindent}

Let us first discuss the parameter $-\lambda$. In principle, one should use its measured value extracted from a fit to the data on the semileptonic inclusive decay $\Lambda_b \to X_c \ell \bar\nu_\ell$, as has been done in the case of mesons. However, to our knowledge, this analysis is not available yet.\
For the time being, following \cite{LS}, we will use a lower bound on $-\lambda$ that can be obtained from data on the heavy baryon spectrum.\par
Let us write down the SR (\ref{3.2e}) for the slope $\rho_\Lambda^2$ and the first and second moments of this SR \cite{LS}, that give the following expressions for $\bar \Lambda$ and $-\lambda$:
\begin{align}
\rho^2_\Lambda &= \sum_{n\geq 0} \left [ \tau_1^{(n)} (1) \right ]^2\;,\label{8.1e}\\
\bar \Lambda &= 2 \sum_{n\geq 0} (\bar \Lambda '^{(n)}-\bar \Lambda) \left [ \tau_1^{(n)} (1) \right ]^2\;,\label{8.2e}\\
- \lambda &= 3 \sum_{n\geq 0}  (\bar \Lambda '^{(n)}-\bar \Lambda)^2  \left [ \tau_1^{(n)} (1) \right ]^2\label{8.3e}\;.
\end{align}
\noindent At leading order in $1/m_Q$, $\bar \Lambda$ and $\bar \Lambda '^{(n)}$ are given by
\begin{align}
\bar \Lambda &= m_{\Lambda_Q(0^+)} - m_Q\label{8.4e}\;,\\
\bar \Lambda '^{(n)} &= m_{\Lambda_Q^{(n)}(1^-)} - m_Q\label{8.5e}\;.
\end{align}

As pointed out in \cite{LS}, Eqns. (\ref{8.2e}) and (\ref{8.3e}) can be combined to give the bound:
\beq
\label{8.6e}
- \lambda \geq {3 \over 2} \bar \Lambda (\bar \Lambda '-\bar \Lambda)
\eeq
where $\bar \Lambda ' = \bar \Lambda '^{(0)}$, i.e. $\bar \Lambda '^{(n)}$ for $n = 0$. The inequality (\ref{8.6e}), that will be used to bound the OPE term in the \emph{r.h.s.} of the inequality (\ref{6.6e}), follows from the fact that $\bar \Lambda '^{(n)}-\bar \Lambda > \bar \Lambda '^{(0)}-\bar \Lambda$ for $n \neq 0$, the $n = 0$ state assumed to be the lowest mass state, a very reasonable hypothesis.\par
In ref. \cite{LS}, $\bar \Lambda$ has been estimated from the masses of the $j^P = 0^+, J^P = {1 \over 2}^+$ states $\Lambda_c (2.286)$, $\Lambda_b(5.620)$, and $\bar \Lambda '-\bar \Lambda$ from the helicity weighted average for the $j^P = 1^-$ doublet $\Lambda_c({1 \over 2}^-)(2.595)$, $\Lambda_c({3 \over 2}^-)(2.625)$. The following results were obtained
\beq
\label{8.7e}
\bar \Lambda \simeq 0.8\ \textrm{GeV} \qquad , \qquad \bar \Lambda '-\bar \Lambda \simeq 0.2\ \textrm{GeV}\;.
\eeq
Then, from the inequality (\ref{8.6e}), one obtains :
\beq
\label{8.8e}
- \lambda \geq 0.24\ \textrm{GeV}^2\;.
\eeq

\section{Theoretical results on $\xi_\Lambda(w)$ and $A(w)$}

\subsection{Heavy Quark Effective Theory}

We have already commented the basic work done within HQET for the exclusive decay $\Lambda_b \to \Lambda_c \ell \overline{\nu}_\ell$. Using the HQET formulation of the heavy hadron fields of arbitrary spin \cite{FALK}, one has recently extended non-forward SR method to the baryon IW function $\xi_{\Lambda}(w)$ \cite{LOR-4}, recovering the lower bound for the slope $\rho_\Lambda^2 = - \xi '_\Lambda (1) \geq 0$ \cite{IWY}, and generalizing it for higher derivatives.
In particular, the new improved bound (\ref{3.4e}) on the curvature $\sigma_\Lambda^2 \geq {3 \over 5}\ [\rho_\Lambda^2+(\rho_\Lambda^2)^2]$  has been derived.\par
With the method based on the Lorentz group \cite{LOR-5}, one has formulated a general criterium to verify if a given phenomenological ansatz for the IW function $\xi_\Lambda(w)$ is consistent with the sum rules, that in turn are equivalent to the Lorentz group approach.\par
Among the various phenomenological results listed in \cite{LOR-5}, it will be useful for our purpose here to recall two possible parametrizations for $\xi_\Lambda(w)$. On the one hand, the ``dipole" form:
\beq
\label{9.1e}
\xi_\Lambda(w) = \left ({2 \over {w+1}} \right)^{2\rho_\Lambda^2}
\eeq

\noi is fully consistent, provided the slope satisfies: 
\beq
\label{9.2e}
\rho_\Lambda^2 \geq {1 \over 4}\;.
\eeq

\noi The curvature is then given by
\beq
\label{9.3e}
\sigma_\Lambda^2 = {\rho_\Lambda^2 \over 2} + (\rho_\Lambda^2)^2\;.
\eeq

On the other hand, the form ($ w = \textrm{ch}(\tau)$): 
\beq
\label{9.4e}
\xi_\Lambda(w) = {\textrm{sh}\left (\tau \sqrt{1-3\rho_\Lambda^2} \right ) \over \textrm{sh}(\tau) \sqrt{1 - 3\rho_\Lambda^2}}
\eeq

\noi is consistent for the whole range of the slope 
\beq
\label{9.5e}
\rho_\Lambda^2 \geq 0
\eeq

\noi allowing saturation of the general lower bound for the slope obtained in \cite{IWY}. In this latter case, the curvature is given by 
\beq
\label{9.6e}
\sigma_\Lambda^2 = {3 \over 5} [\rho_\Lambda^2 + (\rho_\Lambda^2)^2]
\eeq

\noi \emph{i.e.} the lower bound (\ref{3.4e}) is saturated and, according to (\ref{7.5e}), $A'(1)$ vanishes. Of course, this form is an extreme case, constructed in \cite{LOR-5} in order to have saturation of the lower bound.\par

We will later use the form (\ref{9.1e}) for $\xi_\Lambda(w)$ in order to estimated the lower bound on $- \delta_{1/{m_Q^2}}^{(G_1)}$.\par
As far as the subleading form factor $A(w)$ is concerned, HQET implies Eqn. (\ref{1.10bise}) $A(1) = 0$, and the present paper has established two new results, namely, $\sigma_\Lambda^2  = {3 \over 5}[\rho_\Lambda^2+(\rho_\Lambda^2)^2] \Rightarrow A'(1) = 0$ (\ref{7.5e}), and with the Lorentz group approach, $\rho_\Lambda^2  = 0 \Rightarrow A'(1) = 0$ (\ref{7.6e}). 

\subsection{QCD Sum Rules}

As it is well-known, unlike the case of mesons, there are two possible interpolating baryon currents $j_1^v$ and $j_2^v$.\par
The simplest hadronic quantities that have been computed within the QCDSR approach are the $\Lambda_Q$ baryon bound state energy $\bar{\Lambda}$ and the decay constant $f_\Lambda$, that, at leading order, are independent of the choice of the current.\par

\vskip 0.5 truecm

\subsubsection{The value of $\bar{\Lambda}$}

Concerning $\bar{\Lambda}$ for the baryon $\Lambda_Q$, there is consistency between the different calculations. At leading order, one obtains $\bar{\Lambda} = 0.78\ \textrm{GeV}$ in the work of Grozin and Yakovlev \cite{GY}, and $\bar{\Lambda} = (0.79 \pm 0.05)\ \textrm{GeV}$ in the one of Dai \emph{et al.} \cite{DHLLr}, later confirmed in ref. \cite{WHr}. In a series of papers by Groote \emph{et al.} \cite{GKY}, the $\Lambda_b$ baryon mass and decay constant with the radiative corrections at NLO in $\alpha_s$ have been studied, with the finding $\bar{\Lambda} \simeq 0.780\ \textrm{GeV}$.\par

\vskip 0.5 truecm

\subsubsection{The decay constant $f_\Lambda$}

Concerning the decay constant $f_\Lambda$, since the pioneering work of Shuryak \cite{SHURYAK}, the value obtained has somewhat evolved over the years. In ref. \cite{GY}, the range $f_\Lambda= (1.8 - 2.7) \times 10^{-2}\ \textrm{GeV}^3$ was proposed. In the works of Dai \emph{et al.} \cite{DHLLr}\cite{DHHLr}, the value for its square was obtained $f_\Lambda^2 = (0.3 \pm 0.1) \times 10^{-3}\ \textrm{GeV}^6$ (\emph{i.e.} $f_\Lambda \simeq (1.7 \pm 0.3) \times 10^3\ \textrm{GeV}$), confirmed in a much later paper \cite{WHr}. In the more recent work \cite{HJKLr}, a somewhat larger value is found, roughly $f_\Lambda^2 = (1.3 \pm 0.7) \times 10^{-3}\ \textrm{GeV}^6$. In the papers by Groote \emph{et al.} \cite{GKY}, taking into account QCD radiative corrections at NLO, one finds $f_\Lambda \simeq 0.028\ \textrm{GeV}^3$, that is $f_\Lambda^2 \simeq 0.78 \times 10^{-3}\ \textrm{GeV}^6$, i.e. a larger value by roughly a factor of two than that at leading order.

\subsubsection{The leading Isgur-Wise function $\xi_\Lambda(w)$}

Let us shift now to the three-point function, \emph{i.e.} the determination of the IW function $\xi_\Lambda(w)$. Dai \emph{et al.} \cite{DHHLr} performed a QCDSR analysis using both currents ${j}_1^v$ and ${j}_2^v$ and they found for the slope $\rho_\Lambda^2 = 0.55 \pm 0.15$. Huang \emph{et al.} \cite{HJKLr} reanalyzed within the QCDSR the function $\xi_\Lambda(w)$ and found a higher slope, 
$\rho_\Lambda^2 = 1.35 \pm 0.12$.

\subsubsection{Power corrections and $A'(1)$ from QCDSR} 

To get an estimate of the \emph{r.h.s.} of the inequality (\ref{6.6e}), we need the quantity $A'(1)$. Let us now discuss the literature on the subleading form factor $A(w)$.\par
In HQET, as we have seen above, the seminal paper is the analysis of the power corrections in HQET of baryon form factors \cite{FN1r}. The kinetic part of the Lagrangian perturbation at $O(1/m_Q)$ is parametrized by $A(w)$ according to (\ref{1.11e}), that satisfies the condition at zero recoil   $A(1) = 0$ (\ref{1.10bise}). In the present paper, we have seen that there are other rigorous conditions on $A'(1)$ from HQET, 
namely (\ref{7.5e}) and (\ref{7.6e}).\par
We turn to the QCDSR estimate of the function $A(w)$. To our knowledge, there is only the paper by Dai \emph{et al.} \cite{DHHLr} that has computed $A(w)$, given  in terms of a single function $J(w)$ besides the leading IW function $\xi_\Lambda(w)$, with $A(w) \propto J(w)-\xi_\Lambda(w)J(1)$, satisfying then $A(1) = 0$.\par
We find here the following question. We have demonstrated above that within HQET, the slope $A'(1)$ must satisfy the conditions (\ref{7.5e}) and (\ref{7.6e}) that relate the leading and subleading IW functions, at least {\it formally}. Formally, because of course in the real QCD, $\rho_\Lambda^2$ needs not to be zero.\par 
Within the QCDSR approach, according to the generic relation, $A'(1)$ depends on the slope $\rho_{\Lambda}^2$, but it is independent of the curvature $\sigma_{\Lambda}^2$. Therefore, from the two conditions (\ref{7.5e}) and (\ref{7.6e}), we can ask whether the QCDSR calculation satisfies, at least in a formal way, the second condition (\ref{7.6e}), $\rho_\Lambda^2  = 0 \Rightarrow A'(1) = 0$.\par
As far as the case of the current $j_1^v$ is concerned, for $\rho_\Lambda^2  = 0$ all the terms in the corresponding expression for $A'(1)$ are positive-definite and therefore the condition (\ref{7.6e}) cannot be fulfilled. On the contrary, for the case of the current $j_2^v$, for $\rho_\Lambda^2  = 0$ the vanishing of $A'(1)$ is possible, although not in the stability plateau.\par
By the way, notice that the condition $\rho_\Lambda^2  = 0 \Rightarrow A'(1) = 0$ could provide a consistency check for {\it choosing} the right baryonic current, that happens to be $j_2^v$. Interestingly, this constraint seems to lift the ambiguity in the choice of the current.\par
However, choosing the current $j_2^v$ we have obtained lower bounds, depending on the IW function slope, that are large, leading to a lower bound on $- \delta_{1/{m_Q^2}}^{(G_1)} $ that seems to us much too large, as it can for some cases approach 50 \%.

\subsubsection{Summary on QCDSR}

For a number of reasons, we do not proceed any further with the QCDSR method to extract $A'(1)$. First, although the method gives very stable values for $\bar{\Lambda}$, the values obtained for the decay constant $f_\Lambda$ do not show such great stability. Notice that this constant appears also in the determination of the three-point function. Furthermore, as we have seen, $f_\Lambda$ depends strongly on the radiative corrections. On the other hand, we have seen that the study and results of the elastic leading IW function $\xi_\Lambda(w)$ have evolved with time. Concerning $A'(1)$, the quantity of interest to us as far as the bound (\ref{6.6e}) is concerned, there is the paper \cite{DHHLr} that allows to compute it. However, there are subtle points that remain to be investigated, in particular, the fulfillement by the QCDSR approach of the HQET condition, $\rho_\Lambda^2  = 0 \Rightarrow A'(1) = 0$ (\ref{7.6e}).\par
In view of the involvement of the study of the QCDSR method for the purpose of the present paper, we leave it for future work, and turn now to a simpler approach, namely the quark model. As we will see, it allows to compute a value for $A'(1)$ that furthermore satisfies the condition (\ref{7.6e}).

\section{Calculation of $A'(1)$ in the quark model}

\subsection{General formulas}

Let us consider the three four-vectors:
\beq
\label{11.1e}
v = (v^0,0,0,v^z) \qquad , \qquad v' = (v'^0,0,0,v'^z) \qquad \textrm{and} \qquad n = (0,1,0,0)
\eeq
\noi satisfying $v^2 = v'^2 = 1$, $n^2 = -1$ and $v\cdot n=v'\cdot n=0$. Then, from the matrix element (\ref{1.1e}), we obtain:
\begin{align}
\langle\Lambda_c(v',s')\mid\bar{c} {/\hskip - 2 truemm v} b\mid\Lambda_b(v,s)\rangle &= \left(F_1 + F_2 + F_3 w \right) \bar{u}_{\Lambda_c}(v',s')u_{\Lambda_b}(v,s)\;, \label{11.2e}\\
\langle\Lambda_c(v',s')\mid\bar{c} {/\hskip - 2 truemm v'} b\mid\Lambda_b(v,s)\rangle &= \left(F_1 + F_2 w + F_3 \right) \bar{u}_{\Lambda_c}(v',s') u_{\Lambda_b}(v,s)\;,\label{11.3e}\\
\langle\Lambda_c(v',s')\mid\bar{c} {/\hskip - 2 truemm n} b\mid\Lambda_b(v,s)\rangle & = F_1\ \bar{u}_{\Lambda_c}(v',s') {/\hskip - 2 truemm n} u_{\Lambda_b}(v,s)\label{11.4e}
\end{align}
\noi where we have used the constraints ${/\hskip - 2 truemm v}u_{\Lambda_b}(v,s) = u_{\Lambda_b}(v,s)$ and $\bar{u}_{\Lambda_c}(v',s'){/\hskip - 2 truemm v'} = \bar{u}_{\Lambda_c}(v',s')$.\par
It is very convenient to perform the calculation in a particular frame, namely the equal velocity frame (EVF), since in this frame the calculation is symmetric in the exchange $b \leftrightarrow c$. In this frame, one has:
\beq
\label{11.5e}
{\vec q} = {\vec p}_{\Lambda_c}-{\vec p}_{\Lambda_b} \qquad , \qquad {\vec p}_{\Lambda_c} = {m_{\Lambda_c} \over m_{\Lambda_c}+m_{\Lambda_b}}\ {\vec q} \qquad , \qquad {\vec p}_{\Lambda_b} = -{m_{\Lambda_b} \over m_{\Lambda_c}+m_{\Lambda_b}}\ {\vec q}
\eeq

\noi implying $v^z = -v'^z$, $v^0 = v'^0$ and
\beq
\label{11.6e}
{\vec{q}}^{\,2} =(m_{\Lambda_c}+m_{\Lambda_b})^2\frac{(w-1)}{2}\;.
\eeq
In what follows, we take spinors for the bound states as well as for the quarks normalized in the following way:
\beq
\label{11.7e}
u({\vec p},s) = \sqrt{E+m \over 2m}
\left( \begin{array}{c}
\chi \\
{\vec{\sigma}\cdot{\vec p} \over E+m} \chi \end{array} \right)\qquad\textrm{with}\qquad \bar{u}u = 1\;.
\eeq

\noi Of course, the final results are independent of the adopted normalization. We find in the EVF some formulas that will be useful below:
\beq
\label{11.8e}
{E_{\Lambda_c} \over m_{\Lambda_c}} = {E_{\Lambda_b} \over m_{\Lambda_b}} = \sqrt{w+1 \over 2} \qquad \textrm{and}\qquad \bar{u}_{\Lambda_c}(v',s')u_{\Lambda_b}(v,s) = \sqrt{w+1 \over 2} \left( \chi^+_{\Lambda_c} \chi_{\Lambda_b} \right)\;.
\eeq  

\noi From the preceding formulas, we find after some spinor algebra at the bound state level, using (\ref{11.6e}) and particularizing to suitable spin projections:
\begin{align}
F_1(w) =& -\ \sqrt{2 \over w-1} \langle\Lambda_c(v', + \textstyle{{1 \over 2}})\mid\bar{c} {/\hskip - 2 truemm n} b\mid\Lambda_b(v, - \textstyle{{1 \over 2}})\rangle\;,\label{11.9e}\\
F_2(w) =& {1 \over w+1} \left[ \sqrt{2 \over w+1} {1 \over w-1} \langle\Lambda_c(v', + \textstyle{{1 \over 2}})\mid\bar{c} \left(- {/\hskip - 2 truemm v} + w {/\hskip - 2 truemm v'} \right) b\mid\Lambda_b(v, + \textstyle{{1 \over 2}})\rangle \right.\nn\\ 
&\left. +\ \sqrt{2 \over w-1} \langle\Lambda_c(v', + \textstyle{{1 \over 2}})\mid\bar{c} {/\hskip - 2 truemm n} b\mid\Lambda_b(v, - \textstyle{{1 \over 2}})\rangle\right]\;,\label{11.10e}\\
F_3(w) =& {1 \over w+1} \left[ \sqrt{2 \over w+1} {1 \over w-1}  \langle\Lambda_c(v', + \textstyle{{1 \over 2}})\mid\bar{c} \left(w {/\hskip - 2 truemm v} - {/\hskip - 2 truemm v'} \right) b\mid\Lambda_b(v, + \textstyle{{1 \over 2}})\rangle \right.\nn\\ 
&\left. +\ \sqrt{2 \over w-1} \langle\Lambda_c(v', + \textstyle{{1 \over 2}})\mid\bar{c} {/\hskip - 2 truemm n} b\mid\Lambda_b(v, - \textstyle{{1 \over 2}})\rangle\right]\;.\label{11.11e}
\end{align}

As we will check below, after computing within the quark model the matrix elements present in the \emph{r.h.s.} of (\ref{11.9e})-(\ref{11.11e}), the singularities in ${1 \over w-1}$ and in ${1 \over \sqrt{w-1}}$ will disappear.

\subsection{Calculation in the quark model}

We now perform the calculation in the quark model of the form factors $F_i(w)$ ($i = 1, 2, 3$) (\ref{1.3e}), (\ref{1.5e}) and (\ref{1.6e}), together with (\ref{1.7e}) and (\ref{1.8e}). As we will see, the results fulfill this form expected from HQET, and this will allow us to extract $A'(1)$.\par

Let us consider the harmonic-oscillator Hamiltonian for baryons ($i = 1,2, 3$):
\beq
\label{11.12e}
H = \sum_i {{\vec p}_i^{\,2} \over {2 m_i}} + K \sum_{i<j} ({\vec r}_i-{\vec r}_j)^2
\eeq 

\noi where $K$ is independent of flavor. Notice that we use this simple potential because it is the only confining one that allows to separate the center-of-mass in the three-body case of baryons.\par
From now on, we consider the system in which we are here interested, where $m_1 = m_2 = m$ (light quark masses) and $m_3 = m_Q$ (heavy quark mass, $Q = b, c$). For details on what follows on the quark model wave functions, see Appendix A of \cite{JL}. The total and relative coordinates read
\beq
\label{11.13e}
{\vec R} = \displaystyle{{\sum_i m_i {\vec r}_i \over \sum_i m_i}} \qquad , \qquad \rho = {1 \over \sqrt{2}} ({\vec r}_1-{\vec r}_2) \qquad , \qquad \lambda = {1 \over \sqrt{6}} ({\vec r}_1+{\vec r}_2-2{\vec r}_3)\;,
\eeq

\noi and the corresponding conjugate momenta are 
\beq
\label{11.14e}
{\vec p}_R = \sum_i {\vec p}_i \qquad , \qquad {\vec p}_\rho =  {1 \over \sqrt{2}}({\vec p}_1 - {\vec p}_2) \qquad , \qquad {\vec p}_\lambda = \sqrt{3 \over 2} \left[ {m_Q \over \sum_i m_i} ({\vec p}_1 + {\vec p}_2) - {2m \over \sum_i m_i} {\vec p}_3 \right]\;.
\eeq

\noi In terms of these variables the Hamiltonian reads
\beq
\label{11.15e}
H = { {\vec p}_R^{\,2} \over \sum_i m_i} + {{\vec p}_\rho^{\,2} \over 2m} + {{\vec p}_\lambda^{\,2} \over 2\left({3mm_Q \over {2m+m_Q}}\right)} + 3K(\vec{\rho}^{\,2}+\vec{\lambda}^{\,2})
\eeq

In the calculations that follow, we will use the wave functions in momentum space. The Schr$\ddot{\rm{o}}$dinger equation gives, for the ground state, the internal wave function: 
\beq
\label{11.16e}
\psi({\vec p}_\rho,{\vec p}_\lambda) = N_0 \exp \left[-{1 \over 2}({\vec p}^{\,2}_\rho R_\rho^{\,2}+{\vec p}^{\,2}_\lambda R_\lambda^2)\right] 
\eeq 

\noi with 
\beq
\label{11.17e}
 R_\rho^4 = R^4\qquad , \qquad R_\lambda^4 = {2m+m_Q \over 3m_Q} \qquad , \qquad N_0 = \left({3 \sqrt{3} R_\rho^3 R_\lambda^3 \over \pi^3}\right)^{1 \over 2}\;.
\eeq 

We will now calculate in the quark model the matrix elements in (\ref{11.9e})-(\ref{11.11e}):
\begin{align}
\langle\Lambda_c(v', + \textstyle{{1 \over 2}})\mid\bar{c} {/\hskip - 2 truemm v} b\mid\Lambda_b(v, + \textstyle{{1 \over 2}})\rangle &=\langle \Lambda_c(v', + \textstyle{{1 \over 2}})\mid\bar{c} \left(\gamma^0v^0 - \gamma^zv^z \right) b\mid\Lambda_b(v, + \textstyle{{1 \over 2}})\rangle\;,\label{11.18e}\\
\langle\Lambda_c(v', + \textstyle{{1 \over 2}})\mid\bar{c} {/\hskip - 2 truemm v'} b\mid\Lambda_b(v, + \textstyle{{1 \over 2}})\rangle &=\langle\Lambda_c(v', + \textstyle{{1 \over 2}})\mid\bar{c}  \left(\gamma^0v'^0 - \gamma^zv'^z \right) b\mid\Lambda_b(v, + \textstyle{{1 \over 2}})\rangle\;,\label{11.19e}\\
\langle\Lambda_c(v', + \textstyle{{1 \over 2}})\mid\bar{c} {/\hskip - 2 truemm n} b\mid\Lambda_b(v, - \textstyle{{1 \over 2}})\rangle& =-\langle \Lambda_c(v', + \textstyle{{1 \over 2}})\mid\bar{c} \gamma^x b\mid\Lambda_b(v, - \textstyle{{1 \over 2}})\rangle\;.\label{11.20e}
\end{align} 

We need therefore to compute the matrix elements $\langle\Lambda_c(v', + \textstyle{{1 \over 2}})\mid\bar{c} \gamma^0 b\mid\Lambda_b(v, + \textstyle{{1 \over 2}})\rangle$, $\langle\Lambda_c(v', + \textstyle{{1 \over 2}})\mid\bar{c} \gamma^z b\mid\Lambda_b(v, + \textstyle{{1 \over 2}})\rangle$ and
$\langle\Lambda_c(v', + \textstyle{{1 \over 2}})\mid\bar{c} \gamma^x b\mid\Lambda_b(v, - \textstyle{{1 \over 2}})\rangle$. Using the notation $\Gamma = \{\gamma^0, \gamma^z, \gamma^x\}$, we have to compute in the quark model, denoting by 3 the active heavy quark:
\begin{align}
\langle\psi^{(c)}\mid\bar{u}^{(c)}({\vec p}^{\,'}_3) \Gamma u^{(b)}({\vec p}_3)\mid\psi^{(b)}\rangle =&\int {\rm{d}}{\vec p}_1 {\rm{d}}{\vec p}_2 {\rm{d}}{\vec p}_3 {\rm{d}}{\vec p}^{\,'}_1 {\rm{d}}{\vec p}^{\,'}_2 {\rm{d}}{\vec p}^{\,'}_3\ \delta({\vec p}_1-{\vec p}^{\,'}_1)\ \delta({\vec p}_2-{\vec p}^{\,'}_2)\nn\\
&\times\delta \textstyle{\left({\vec p}^{\,'}_1+{\vec p}^{\,'}_2+{\vec p}^{\,'}_3- {m_{\Lambda_c} \over m_{\Lambda_c}+m_{\Lambda_b}} {\vec q}\right)}\ \delta \textstyle{\left({\vec p}_1+{\vec p}_2+{\vec p}_3+ {m_{\Lambda_b} \over m_{\Lambda_c}+m_{\Lambda_b}} {\vec q} \right)}\nn\\
&\times\psi^{(c)\,+}\left({\vec p}^{\,'}_1,{\vec p}^{\,'}_2,{\vec p}^{\,'}_3 \right)\bar{u}^{(c)}({\vec p}^{\,'}_3) \Gamma u^{(b)}({\vec p}_3)\ \psi^{(b)} \left({\vec p}_1,{\vec p}_2,{\vec p}_3 \right)\;.\label{11.21e}
\end{align}
\noi We obtain, after some algebra and change of variables:
$$\langle\psi^{(c)} \mid \bar{u}^{(c)}({\vec p}^{\,'}_3) \Gamma u^{(b)}({\vec p}_3) \mid \psi^{(b)}\rangle = {1 \over 3\sqrt{3}} \int {\rm{d}}{\vec{p}}_\rho {\rm{d}}{\vec{p}}_\lambda \psi^{(c)\,+}\textstyle{\left({\vec{p}}_\rho,{\vec{p}}_\lambda - \sqrt{{3 \over 2}} {2m \over 2m+m_c} {m_{\Lambda_c} \over m_{\Lambda_c}+m_{\Lambda_b}} {\vec{q}} \right)}$$
$$\times\bar{u}^{(c)}\textstyle{\left(- \sqrt{{2 \over 3}} {\vec p}_\lambda + {m_{\Lambda_c} \over m_{\Lambda_c}+m_{\Lambda_b}} {\vec q} \right)} \Gamma u^{(b)}\textstyle{\left(- \sqrt{{2 \over 3}} {\vec p}_\lambda - {m_{\Lambda_b} \over m_{\Lambda_c}+m_{\Lambda_b}} {\vec q} \right)}\ \psi^{(b)}\textstyle{\left({\vec p}_\rho,{\vec p}_\lambda + \sqrt{{3 \over 2}} {2m \over 2m+m_b} {m_{\Lambda_b} \over m_{\Lambda_c}+m_{\Lambda_b}} {\vec q} \right)}\;.$$
\beq
\label{11.22e}
\eeq

Some important words of caution are in order here concerning the terms that we keep in the non-relativistic expansion. Following the usual methods in the quark model, we expand the quark spinor matrix element in (\ref{11.22e})
\beq
\label{11.23e}
\bar{u}^{(c)}\textstyle{\left(- \sqrt{{2 \over 3}} {\vec p}_\lambda + {m_{\Lambda_c} \over m_{\Lambda_c}+m_{\Lambda_b}} {\vec q} \right)} \Gamma u^{(b)}\textstyle{\left(- \sqrt{{2 \over 3}} {\vec p}_\lambda - {m_{\Lambda_b} \over m_{\Lambda_c}+m_{\Lambda_b}} {\vec q} \right)}
\eeq
\noi up to first order in $1/m_Q$ and then we perform the integral on the internal variables.\par
On the other hand, we keep terms of the order $R^2m^2$, that are of order $\left({v \over c}\right)^{-2}$, corresponding to the non-relativistic limit of the slope of the IW function (for a discussion, see \cite{MLOPR}). An important remark to be made also is that the HQET parameter $\bar{\Lambda}$ can be decomposed, in the quark model, under the form $\bar{\Lambda} = 2m + \epsilon$, where $m$ is the constituent light quark mass and $\epsilon$ is the binding energy in the non-relativistic quark model. Of course, both cannot be distinguished in HQET. However, in a non-relativistic expansion as adopted here, $\epsilon$ is of order ${1 \over R^2m}$ that, relatively to the constituent mass $m$ is a subleading quantity ${\epsilon \over m} \sim \left({v \over c}\right)^2$. Therefore, from now on, we will neglect $\epsilon$ and take $2m = \bar{\Lambda}$.\par 
On the other hand, to extract $A'(1)$ we need to go to the orders $w-1$,  ${\bar{\Lambda} \over m_Q}$ and ${\bar{\Lambda} \over m_Q}(w-1)$, that we will also keep in our calculation.\par
To summarize, in our quark model calculation we keep the following orders in the non-relativistic expansion: leading order in HQET $\left(\frac{\bar{\Lambda}}{m_Q}\right)^0$ (keeping the order $\bar{\Lambda}^2R^2$ and the first order in $w-1$), order ${\bar{\Lambda} \over m_Q}$ and order ${\bar{\Lambda} \over m_Q}(w-1)$.\par

A lengthy and careful calculation gives the following results:
\begin{align}
\langle\Lambda_c(v', + \textstyle{{1 \over 2}})\mid\bar{c} \gamma^0 b\mid\Lambda_b(v, + \textstyle{{1 \over 2}})\rangle & \cong F(w)\;,\label{11.24e}\\
\langle\Lambda_c(v', + \textstyle{{1 \over 2}})\mid\bar{c} \gamma^z b\mid\Lambda_b(v, + \textstyle{{1 \over 2}})\rangle & \cong -\ F(w) \sqrt{w-1 \over 2} \left({\bar{\Lambda} \over 2m_b} - {\bar{\Lambda} \over 2m_c} \right) \left(1 - {w-1 \over 4} \right)\;,\label{11.25e}\\
\langle\Lambda_c(v', + \textstyle{{1 \over 2}})\mid\bar{c} \gamma^x b\mid\Lambda_b(v, - \textstyle{{1 \over 2}})\rangle & \cong F(w) \sqrt{w-1 \over 2}\left(1 + {\bar{\Lambda} \over 2m_b} + {\bar{\Lambda} \over 2m_c} \right)\;.\label{11.26e}
\end{align}
\noi where 
\beq
\label{11.27e}
F(w) = \exp \left[- {\sqrt{3} \over 4}\ \bar{\Lambda}^2R^2 \left(1 + {\bar{\Lambda} \over 4m_b} + {\bar{\Lambda} \over 4m_c} \right) (w-1)\right]
\eeq
\noi \emph{i.e.} we obtain in the heavy quark limit the IW function $F(w) \to \xi(w)$ and its slope $\rho^2$:
\beq
\label{11.28e}
\xi_{\Lambda}(w) = \exp \left[- {\sqrt{3} \over 4}\ \bar{\Lambda}^2R^2 (w-1)\right] \qquad \textrm{with} \qquad \rho^2_{\Lambda} =  {\sqrt{3} \over 4}\ \bar{\Lambda}^2R^2\;.
\eeq

One has demonstrated elsewhere \cite{LOR-5} that, strictly speaking, the exponential form (\ref{11.28e}) is not consistent with Lorentz group constraints. However, not unexpectedly in the non-relativistic harmonic-oscillator quark model, one has such a behaviour in the EVF. However, this is not really an inconvenience since the non-relativistic expansion cannot hold at large $w-1$, but only near zero recoil, in the neighborhood of $w = 1$. Therefore, we will only consider the first values in the Taylor expansion of $\xi_{\Lambda}(w)$ in (\ref{11.28e}), namely the normalization $\xi_{\Lambda}(1) = 1$ and the slope $-\xi'_{\Lambda}(1) \equiv \rho_{\Lambda}^2$. Interestingly, the value $R^2 \simeq 6\ \rm{GeV}^{-2}$ extracted from the light baryon spectrum \cite{JL} gives numerically a value for the slope that is not non-sense ($\bar{\Lambda} \simeq 2m \simeq 0.70\ \rm{GeV}$): 
\beq
\label{11.28-1e}
\rho_{\Lambda}^2 \simeq 1.27\;. 
\eeq

We now go back to the calculation. Taking into account that, from (\ref{11.5e}) and (\ref{11.6e}): 
\beq
\label{11.29e}
v^z = - v'^z = - \sqrt{{w-1 \over 2}} \qquad \textrm{and} \qquad \qquad v^0 = v'^0 = \sqrt{{w+1 \over 2}}\;,
\eeq
\noi we can now compute (\ref{11.18e})-(\ref{11.20e}):
\begin{align}
\langle\Lambda_c(v', + \textstyle{{1 \over 2}})\mid\bar{c} {/\hskip - 2 truemm v} b\mid\Lambda_b(v, + \textstyle{{1 \over 2}})\rangle &= F(w)\left[\sqrt{{w+1 \over 2}} - {w-1 \over 2}\left(\frac{\bar{\Lambda}}{2m_b} - \frac{\bar{\Lambda}}{2m_c} \right) \left(1 - {w-1 \over 4} \right) \right]\;,\label{11.30e}\\
\langle\Lambda_c(v', + \textstyle{{1 \over 2}})\mid\bar{c} {/\hskip - 2 truemm v'} b\mid\Lambda_b(v, + \textstyle{{1 \over 2}})\rangle &= F(w)\left[\sqrt{{w+1 \over 2}} + {w-1 \over 2}\left(\frac{\bar{\Lambda}}{2m_b} - \frac{\bar{\Lambda}}{2m_c} \right) \left(1 - {w-1 \over 4} \right) \right]\;,\label{11.31e}\\
\langle\Lambda_c(v', + \textstyle{{1 \over 2}})\mid\bar{c} {/\hskip - 2 truemm n} b\mid\Lambda_b(v, - \textstyle{{1 \over 2}})\rangle &= - F(w) \sqrt{w-1 \over 2}\left(1 + {\bar{\Lambda} \over 2m_b} + {\bar{\Lambda} \over 2m_c} \right)\;.\label{11.32e}
\end{align}

Inserting these expressions in the formulas (\ref{11.9e})-(\ref{11.11e}), we obtain:
\begin{align}
F_1(w) &= F(w)\left(1 + {\bar{\Lambda} \over 2m_b} + {\bar{\Lambda} \over 2m_c} \right)\;,\label{11.33e}\\
F_2(w) &=  {F(w) \over w+1} \left[\sqrt{{w+1 \over 2}}\left(\frac{\bar{\Lambda}}{2m_b} - \frac{\bar{\Lambda}}{2m_c} \right) \left(1 - {w-1 \over 4} \right) - \left(\frac{\bar{\Lambda}}{2m_b} + \frac{\bar{\Lambda}}{2m_c} \right) \right]\;,\label{11.34e}\\
F_3(w) &= {F(w) \over w+1} \left[\sqrt{{w+1 \over 2}}\left(\frac{\bar{\Lambda}}{2m_c} - \frac{\bar{\Lambda}}{2m_b} \right) \left(1 - {w-1 \over 4} \right) - \left(\frac{\bar{\Lambda}}{2m_b} + \frac{\bar{\Lambda}}{2m_c} \right) \right]\;.\label{11.35e}
\end{align}
It is important to point out that the apparent singularities in the expressions  (\ref{11.9e})-(\ref{11.11e}) have disappeared, since the singularities at the bound state level are cancelled by their inverses at the quark level. Notice the different role of $m_b$ and $m_c$ in the first term of the expressions for $F_2(w)$ (\ref{11.34e}) and $F_3(w)$ (\ref{11.35e}).

\subsection{Final results and comparison with HQET}

From the expressions (\ref{11.33e})-(\ref{11.35e}), we can compute the quantities $F_i(1)$ ($i = 1, 2, 3$):
\begin{align}
F_1(1) &= 1 + {\bar{\Lambda} \over 2m_b} + {\bar{\Lambda} \over 2m_c}\;,\label{11.36e}\\
F_2(1) &= - {\bar{\Lambda} \over 2m_c}\;,\label{11.37e}\\
F_3(1) &= - {\bar{\Lambda} \over 2m_b}\;.\label{11.38e}
\end{align}
\noi and the derivatives $F'_i(1)$ ($i = 1, 2, 3$): 
\begin{align}
F'_1(1) &= - \rho^2\left[1 + {3 \over 2}\left({\bar{\Lambda} \over 2m_b} + {\bar{\Lambda} \over 2m_c}\right) \right]\;,\label{11.39e} \\ 
F'_2(1) &= {1 \over 2} {\bar{\Lambda} \over 2m_c} + \rho^2 {\bar{\Lambda} \over 2m_c}\;,\label{11.40e}\\
F'_3(1) &= {1 \over 2} {\bar{\Lambda} \over 2m_b} + \rho^2 {\bar{\Lambda} \over 2m_b}\;,\label{11.41e}
\end{align}

Let us now compare with the HQET formulation, Eqns. (\ref{1.3e}), (\ref{1.5e}) and (\ref{1.6e}). We first realize that there is consistency between the general structure in ${1 \over m_Q}$ of $F_2(w)$ and $F_3(w)$ and the quark model results.\par

In the quark model, the function $B_2(w)$ (\ref{1.8e}) satisfies:
\beq
\label{11.42e}
B_2(1) = - \bar{\Lambda} \qquad \textrm{and} \qquad \qquad B'_2(1) = {\bar{\Lambda} \over 2} + \rho^2 {\bar{\Lambda}}\;.
\eeq

Notice the very non-trivial point that $B_2(1)$, the derivative $B'_2(1)$ (\ref{11.42e}) as well as $F_1(1)$ (\ref{11.36e}) \emph{coincide exactly with the HQET results}, as can be read from the general formulas (\ref{1.3e}), (\ref{1.7e}) and (\ref{1.8e}).\par
 
Let us now compare the first derivative $F'_1(1)$ between the quark model and HQET. In HQET, one has from (\ref{1.3e}):
\beq
\label{11.43e}
F'_1(1) = -\rho^2 + \left({1 \over 2m_b} + {1 \over 2m_c}\right)\left[B'_1(1)-B'_2(1)\right]
\eeq

\noi that can be expressed using (\ref{1.7e}) in the form
\beq
\label{11.44e}
F'_1(1) = -\rho^2 + \left({1 \over 2m_b} + {1 \over 2m_c}\right)\left[{\bar{\Lambda} \over 2} + A'(1)-B'_2(1)\right]
\eeq

\noi with $B'_2(1)$ given by (\ref{11.42e}). We now compare with the quark model result (\ref{11.39e}). This yields the equality:
\beq
\label{11.45e}
- \rho^2 + \left({1 \over 2m_b} + {1 \over 2m_c}\right)\left[{\bar{\Lambda} \over 2} + A'(1)-{\bar{\Lambda} \over 2} - \rho^2\bar{\Lambda}\right] =  - \rho^2\left[1 + {3 \over 2}\left({\bar{\Lambda} \over 2m_b} + {\bar{\Lambda} \over 2m_c}\right)\right]
\eeq

\noi from which can be extracted the expression for $A'(1)$:
\beq
\label{11.46e}
A'(1) = - {\bar{\Lambda} \over 2} \rho^2
\eeq

\noi which satisfies the constraint $A'(1) \to 0$ for  $\rho^2 \to 0$ and which is the main result of this Section.

\section{Lower bound on $- \delta_{1/{m_Q^2}}^{(G_1)}$}

It might be enlightening to split the inequality (\ref{6.6e}) into two individual contributions, one from OPE, the other from the sum of intermediate states (IS), respectively: 
\bea
\label{12.1e}
- \delta_{1/{m_Q^2}}^{(G_1)} \geq \Delta_{OPE} + \Delta_{IS}
\eea

\noi where
\bea
\label{12.2e} 
\Delta_{OPE} = - {1 \over 2} \left [ \left ({1 \over 2m_c}  - {1 \over 2m_b} \right )^2 + {8 \over 3} {1 \over 2m_c} {1 \over 2m_b} \right ] \lambda
\eea
and
\bea
\label{12.3e} 
\Delta_{IS} = {3 \over 10} \left ( {1 \over 2m_c} - {1 \over 2m_b}\right )^2 {[A'(1)]^2 \over \sigma_\Lambda^2 - {3 \over 5}[\rho_\Lambda^2+(\rho_\Lambda^2)^2]}\;.
\eea

With the masses $m_c = 1.1\ \textrm{GeV}$, $m_b = 4.5\ \textrm{GeV}$ and the bound $- \lambda \geq 0.24\ \textrm{GeV}^2$ (\ref{8.8e}), one gets: 
\bea
\label{12.5e}
\Delta_{OPE} \geq 0.03\;.
\eea

To bound $\Delta_{IS}$, one needs not only information on $A'(1)$ but also the shape of the IW function to have the slope $\rho_\Lambda^2$ and the curvature $\sigma_\Lambda^2$.\par
For the leading IW function, we adopt here the ``dipole" shape (\ref{9.1e}), that implies (\ref{9.3e}).
For the quantity $A'(1)$, we adopt the expression (\ref{11.46e}) obtained in the quark model. We use the reasonable value $\bar{\Lambda} = 0.7\ \rm{GeV}$, twice the light quark constituent mass, that agrees qualitatively with the QCDSR calculations, as we have seen in Section 10.2.1. Then, $\Delta_{IS}$ depends only on the slope $\rho^2$, that we vary. We find:
\begin{align}
\textrm{for}\;\;\rho_\Lambda^2 = 0.5\;\;,\;\;- \delta_{1/{m_Q^2}}^{(G_1)} \geq 0.052\\
\textrm{for}\;\;\rho_\Lambda^2 = 1.0\;\;,\;\;- \delta_{1/{m_Q^2}}^{(G_1)} \geq 0.045\\
\textrm{for}\;\;\rho_\Lambda^2 = 1.5\;\;,\;\;- \delta_{1/{m_Q^2}}^{(G_1)} \geq 0.043\label{12.6e}
\end{align}

Of course, one must keep in mind that the contribution of the $1^+$ states in the sum (\ref{5.4e}) has not been estimated, only their positivity is taken into account. This suggests that the actual value of $- \delta_{1/{m_Q^2}}^{(G_1)}$ could be relatively large. Another remark to be made is that in the baryon case, there is no strong cancellation like for the bound that appeared in the meson case due to the various subleading form factors $\chi_i (w)$ (i = 1,2,3) \cite{JLOR-1}. A single form factor $A(w)$ contributes in the present case.\par
To conclude, one gets a relatively large lower bound for $- \delta_{1/{m_Q^2}}^{(G_1)}$, although, of course, the final numerical results rely on two phenomenological hypotheses: the "dipole" shape (\ref{9.1e}) for $\xi(w)$ and the quark model estimate (\ref{11.46e}) for $A'(1)$.\par
It is quite important to have an estimation of $A'(1)$ in the QCDSR approach. As stated above, one needs to clarify in detail some important points of this theoretical method, and we plan to do it in near future.\par 
The crucial decisive next step would be to have a measurement on the lattice of the correction at zero recoil $- \delta_{1/{m_Q^2}}^{(G_1)}$, as has been done already for the meson case.

\section{Conclusions}

We have obtained new results in the heavy quark expansion of Heavy Quark Effective Theory that are relevant for the differential rate of the decay
$\Lambda_b \to \Lambda_c \ell \bar{\nu}_\ell$, that will be measured in greater precision at LHCb. We have written down a SR for the elastic subleading form factor $A(w)$ at order $1/m_Q$, that originates from the Lagrangian perturbation $\mathcal{L}_{kin}$. This result, together with another SR in the forward direction for the axial-vector form factor $|G_1(1)|^2$, has allowed us to obtain a lower bound for the correction at zero recoil $- \delta_{1/{m_Q^2}}^{(G_1)}$ in terms of the derivative $A'_1(1)$ and the slope $\rho^2_\Lambda$ and curvature $\sigma^2_\Lambda$ of the leading Isgur-Wise function $\xi_{\Lambda}(w)$. We have found that the derivative $A'(1)$ must vanish in the formal limits $\sigma^2_\Lambda \to {3 \over 5}[(\rho^2_\Lambda)^2+\rho^2_\Lambda]$ and $\rho^2_\Lambda \to 0$, establishing a non-trivial correlation between the shape of the leading IW function $\xi_{\Lambda}(w)$ and the subleading one $A(w)$. We have discussed results of the theory for the leading and subleading IW functions (HQET and QCDSR). Moreover, we have performed a calculation in the quark model of the leading and subleading form factors, that agree with HQET up to order and including $(w-1)$. Consistently, the result for $A'_1(1)$ in the quark model vanishes in the limit $\rho^2_\Lambda \to 0$.  We finally bound from below the $1/m_Q^2$ correction $- \delta_{1/{m_Q^2}}^{(G_1)}$ to the axial form factor at $w = 1$, that contributes to the differential rate of $\Lambda_b \to \Lambda_c \ell \bar{\nu}_\ell$ at zero recoil.

\section*{Acknowledgments} \hspace*{\parindent} The work of F. Jugeau has been supported by the National Natural Science Foundation of China (NSCF) under grant No. Y01161005Z and is presently supporterd by the Brazilian National Counsel of Technological and Scientific Development (CNPq)  under the fellowship No. 150252/2011-0. The research of Yu Jia is supported in part by the National Natural Science Foundation of China under Grant No. 10875130, 10935012. One of us (L. Oliver) is indebted to the Theoretical Physics Division of the IHEP (Beijing) for support during the realisation of this work. We are also indebted to Professor Chun Liu for useful information on the QCDSR calculations of leading and subleading IW functions, to Alain Le Yaouanc for an interesting remark on this topic, and to Professor Adam Leibovich for very useful correspondence.

\vskip 1 truecm

\end{document}